# Giant Magnetic-Field-Induced Strain in $Ni_2MnGa$-based polycrystal

A. A. Mendonça[1], J. F. Jurado[2], S. J. Stuard[1], L. E.L. Silva[1], G. G. Eslava[1], L. F. Cohen[3], L. Ghivelder[1] and A. M. Gomes[1]

1. Instituto de Física, Universidade Federal do Rio de Janeiro, C.P. 68528, Rio de Janeiro, 21941-972, Brazil.

2. Lab. Propiedades Térmicas Dieléctricas de Compositos, Depto. de Física y Química, Universidad Nacional de Colombia, A.A 127, Manizales, Colombia.

3. Blackett Laboratory, Imperial College London, Prince Consort Rd., London, United Kingdom SW7 2BZ.

## Abstract

Ferromagnetic $Ni_2MnGa$-based alloys play an important role in technological fields, such as smart actuators, magnetic refrigeration and robotics. The possibility of obtaining large non-contact deformation induced by an external perturbation is one of its key strengths for applications. However, the search for materials with low cost, practical fabrication procedures and large signal output under small perturbing fields still poses challenges. In the present study we demonstrate that by judicial choice of substitution on the Mn site, an abrupt magnetostructural transition from a paramagnetic austenite phase to a ferromagnetic martensite one can be tuned to close to room temperature achieving large and reproducible strains. The required magnetic field to induce the strain varies from small values, as low as 0.25 T for 297.4 K and 1.6% of strain, to 8 T for 305 K and 2.6% of strain. Our findings point to encouraging possibilities for application of shape memory alloys in relatively inexpensive, scalable polycrystalline materials.

# 1. Introduction

The possibility to perform large shape deformation in an alloy without contact is highly attractive for applications, since it avoids mechanical wear and energy losses by frictional forces. Materials with magnetic-field-induced (MFI) strain are named as ferromagnetic shape memory alloys (FSMA) which have two main mechanisms to generate large deformations: martensitic direct-reverse transformation, also called metamagnetic shape memory effect, [1-4] and reorientation of the martensite variants by twin boundary motion.[5-8] Although samples with a twin boundary motion mechanism can present large values of strain under magnetic fields smaller than 1 T, usually their fabrication process is unattractive for large scale production because the strain effect is only significant in single crystals or in polycrystalline alloys where pores are smaller than the grain size are introduced. [9-11] On the other hand, the metamagnetic shape memory effect can take place by application of various annealing processes creating MFI deformation in polycrystalline alloys supporting martensitic direct-reverse transformations. Nonetheless, even in this case, it is hard to find materials with large transformation strain induced by small magnetic fields, since the transition does not tend to be sufficiently abrupt. The large thermal hysteresis of these compounds, typically higher than 10 K, also promotes difficulties for its use.[12]

$Ni_2MnGa$-based Heusler alloys are extensively studied because they present a ferromagnetic shape memory effect. In particular, the $Ni_2MnGa$ compound crystallizes in a cubic austenite $L2_1$-type structure (space group Fm-3m) with a second order ferromagnetic to paramagnetic transition at 376 K and martensitic transition at approximately 200 K.[13,14] Often, when an atom with more valence electrons than some precursor of the $Ni_2MnGa$-based alloys is introduced in the material and small changes of lattice parameters or electron hybridization occur, the result is a increase of the martensitic transition temperature.[15,16,17] In contrast, the Mn replacement by non-ferromagnetic elements decreases the transition temperature as a result of weakening the Mn-Mn ferromagnetic exchange interactions.[15-18] As an example, Mn replacement by Cu leads to an approximate coincident magnetostructural transition (MST), at around 308K.[18]

Usually, the MST is an important property for FSMA due to a large difference of magnetic susceptibility between the martensite and austenite phases. For a sample exposed to a magnetic field in the paramagnetic phase, the supplied energy is used to align the ferromagnetic phase and as a consequence, both magnetic and structural transitions can occur simultaneously, once they are coupled, as illustrated in Figure 1.

In order to reduce the cost of production of these alloys, it is advisable to replace Ga by cheaper elements, such as Al, as suggested by Mejia et al.[19] The Al insertion in $Ni_2MnGa$ generates a coexistence of $L_{21}$ (ferromagnetic) and $B_2$ (antiferromagnetic) structures, leading to a predominant antiferromagnetism when $x > 0.3$.[20,21] Tuning the Cu composition also reveals important properties for application. For example we have previously shown that the $x = 0.2$ compound ($Ni_2Mn_{0.8}Cu_{0.2}Ga_{0.9}Al_{0.1}$), shows sizable magnetocaloric properties around 295 K,[19] and in the present study, we demonstrate that the $Ni_2Mn_{0.7}Cu_{0.3}Ga_{0.84}Al_{0.16}$ alloy presents an unusually large MFI strain under low magnetic fields at room temperature due to its sharp MST.

## 2. Materials and Methods

A bulk of 1.5 g with composition $Ni_2Mn_{0.7}Cu_{0.3}Ga_{0.84}Al_{0.16}$ was fabricated using conventional arc melting in 99.999% pure argon atmosphere and metallic elements of purity at least 99.99%. The samples were re-melted 3 times with care not to keep the arc on more than five seconds, and therefore avoiding large losses due to Mn vaporization. Initially, Mn losses of approximately 3% at the end of the melting process were observed, and to account for this, we added 3% excess of Mn before the melting to ensure correct stoichiometry at the end. To achieve greater homogenization, two thermal treatments were applied. To accomplish this, the sample was wrapped with tantalum foil and encapsulated in quartz tubes under a low argon pressure of 0.2 atm. The first thermal treatment was done for 72 h at 1273 K and the second for 24 h at 673 K, both using a rate of 3 K/min and quenching in room temperature water at the end of each process. For the measurements, the pieces of sample were sawed by a diamond blade. The identification of the crystal structure was made with a Panalytical X'Pert Pro powder X-ray diffractometer using Cu Kα radiation. An investigation of the microstructure was performed in the Leica DMRM optical microscope. The dilatometry as a function of temperature as well as the magnetostriction measurements were made in the silver based Capacitance Dilatometer DIL20-11,[22] placed in the PPMS platform. For these measurements, the temperature and the magnetic field rates were 0.2 K/min and 3 mT/s, respectively. Isothermal and isofield magnetization measurements using a Vibrating Sample Magnetometer (VSM). We used magnetic field up to 9 T with a ramp of 5 mT/sec. Vickers hardness measurements with a applied load of 50 g were made in a Leitz Wetzlar micro-hardness tester microscope at room temperature, and the data presented is an average value of four indentations.

# 3. Results and Discussion

## 3.1 Structural properties

X-ray powder diffraction (XPD) measurements are shown in the Figure 2a. The sample was initially measured at 170 K, and subsequently heated to the next temperatures, 217 K, 257 K and 298 K. The diffractogram at 298 K exhibits the coexistence of the austenitic $L2_1$-type cubic structure (space group Fm-3m) and the martensitic $L1_0$ non-modulated tetragonal structure achieved when the sample is cooled below its MST temperature. The intermartensitic transition is not observed due to the non-modulated tetragonal structure, which is the ground state for martensitic phases. In general, $Ni_2MnGa$-based Heusler alloys may present intermartensitic transitions among two modulated martensitic structures (five layered 5M to seven layered 7M) or a transformation from a modulated (5M or 7M) to a non-modulated structure.[23] Therefore, the direct transformation from the austenite to a non-modulated martensite phase yields a particular advantage in that  the structure does not change again. Furthermore, through Rietveld refinement we observe that, as the temperature rises, the tetragonal parameters of the unit cell $a = b$ increase while the $c$ parameter decreases, as shown in the Figure 2b. In addition, the unit cell volume grows, as noticed in the inset of Fig. 2b, and the tetragonality relation *c*/*a*', which describes the unit cell symmetry, decreases (Table 1). From the refinement of the data at 298 K, we calculated that around 16% of the sample is in a cubic parent phase with  lattice parameter 5.829 Å, which leads to a unit cell volume of 198.05 Å$^3$.

In Figure 2c optical micrographs of the polished material are presented. In parts I and II the surface regions without a plate-like pattern represent the austenite phase while the regions with martensitic phase exhibits a twinned structure which enable us to conclude for that phase coexistence region to exist.  What causes twins plates in the martensite phase is a shear-like process in the stage of nucleation and structural growth. This formation of twinned structure in martensite phase was observed before in other alloys with paramagnetic austenite and ferromagneticc martensite.[15,24] In parts III and IV the sample presents a region were the martensite phase was established. One can notice that in the phase coexistence regime that the width of the plates decreases whenever the sample structure is closer to the austenite phase. The widths of the martensitic twin plates observed at room temperature varied from 1 μm to 30 μm.

We also measured the Vickers hardness of the sample at room temperature. The average hardness is (268 $\pm$ 8) HV (50 g), which is a value similar to other reported polycrystalline $Ni_2MnGa$ compounds.[25] By optical microscopy, no cracks were found in the region of the indentations. This indicates a small brittleness in the material, even with the two new elements added to the parent $Ni_2MnGa$-based alloy.

3.2 Magnetization and Thermal Expansion

Field-cooled-cooling (FCC) magnetization measurements are presented in Figure 3. The material shows a transformation from the paramagnetic austenitic phase to the ferromagnetic martensitic one, and this MST occurs at approximately 1 $KT^{-1}$ rate. Usually, among common FSMA, a difference of at least 5 K is observed between the start temperature ($M_S$) and the final temperature ($M_F$) of the martensitic transition,[7,9,12] while in our case it is approximately 2 K. In general, a weak magnetic field dependence of the transformation temperature is a drawback for applications because it implies the need for a large magnetic field to perform the phase change for MFI applications. However if the material presents an abrupt transition, a low magnetic field is sufficient to completely induce the transition, because the total temperature interval over which the transformation is confined, is small. Therefore, an abrupt transition compensates for the low magnetic field dependence of the transformation temperature.

Dilatometry measurements, plotted in Figure 4, show a relative thermal expansion of 2.6%, that indicates the presence of a large strain in the compound. This is an impressive value for a polycrystalline material.[8,27] The same field rate dependence of the transition temperature was also observed in the dilatometry when the field was fixed at 5 T. A moderate thermal hysteresis of 8 K was calculated as the difference $A_F$ - $M_S$, where $A_F$ is the final temperature of the austenitic transition. The value $M_S$ - $M_F$ $\approx$ 2 K is in agreement with the magnetization data.

For ferromagnetic shape memory alloys, an abrupt magneto-structural change is interesting because it requires a lower change of the external stimulus (temperature or magnetic field) to produce a considerable signal output. Under high magnetic fields the sample presents the same value of deformation as at zero field. This phenomenon could indicates that the large deformation is due to the volume difference between the phases. However, a simple calculation using the lattice parameters show that the maximum volume difference between the phases is around 5.5% (the cubic cells have twice as many atoms as tetragonal cells), and experimentally we found a one-dimensional strain of 2.6%. Therefore, some effect due to alignment of variant of martensite must be considered. Although the material studied is polycrystalline, it can present some grains as large as 0.5 mm (Fig 2b part III and IV were achieved inside two different grains). We used a small peace of sample for the dilatometric measurements, with size on the order of 2 mm, then, it is possible that the small amount of grain boundaries results in negligible constraints to obtain large deformation in the transition between the cubic and the oriented martensite variants. Since transformation stresses also can result in texture of martensite variants, it is possible to measure the strain in the absence of magnetic field as well.

In addition, at zero applied magnetic field, as well as at 5 T, the onset of the MST on cooling is smooth, but suddenly an abrupt transformation takes place. From this behavior, it was possible to identify a temperature region where the compound exhibits phase coexistence. In each phase, the coexistence rises until a critical value, where the largest structural change occurs in a small temperature interval. The avalanche-like process is characterized by an accelerated increase of the martensitic microstructure under continuous temperature change. The nucleation of the low temperature phase is achieved at the temperature at which the available energy to change the phase is enough to overcome the energy barrier of the transformation. That available energy originates from atomic interactions, mainly by the magnetism of the 3d orbitals from Ni, Mn and Cu which are 3d metals. Since the compound is cooled in the paramagnetic austenite phase, the atoms from the cubic structure get closer until the magnetic interactions (mainly the Mn-Mn distances) are strong enough to activate the transformation and generate the nucleation of the asymmetric tetragonal phase from the symmetric cubic structure. When the nucleation occurs in the MST, the structural growth of the martensite phase is accompanied by an increase of the exchange interactions due to the formation of martensitic ferromagnetic domains in the phase coexistence region. Because of that, although the constant temperature change sweep procedure leads to the gradual appearing of the martensite phase in the MST, the interplay between structure growth and ferromagnetism contribute to complete the transition as a run away avalanche-like process.

## 3.3 Magnetostructural dynamics in the phase coexistence region

In order to study the evolution of structure and ferromagnetism in the MST, we performed magnetization and magnetostriction measurements as a function of magnetic field at 298 K, as shown in Figure 5. At this temperature, the sample presents phase coexistence with predominance of the austenitic phase (see Figure 4). We observe clearly in Figure 5 that an abrupt magnetization change comes suddenly after the sharp structural change. This is due to the structural dependence of the ferromagnetism, since only the low temperature phase is ferromagnetic. Quantatively, the critical field of the magnetostriction measurement is (0.69 $\pm$ 0.01) T while the critical field of the magnetization curve is (0.75 $\pm$ 0.01) T.

Before the sharp structural change, the material already presents a partial mix between the austenite and martensite phases. This leads to the existence of ferromagnetic ordering at that stage, due to the presence of the martensite phase in the phase coexistence. As a consequence, while performing the transition, the strong magnetic interactions from the product phase help to accelerate the development of the low temperature structure, and in this sense the positive feedback from the product phase leads to an overall sharper transformation.

### 3.4 Magnetic-field-induced deformation

Magnetostriction measurements without compressive pre-strain were made and show a contraction while performing the phase transition under the magnetic field change, as seen in Figure 6. Before each measurement, the sample was heated up to the austenite phase and then it was cooled to the target temperature. From its temperature dependence (see Figure 4), it is noticeable that, for each isotherm measured in the cubic phase, the initial size of the sample was not the same because of the phase coexistence state. However, at the end of the process, the material goes to the same size in all measured isotherms, due to the full induction of the martensite phase by the magnetic field. This way, in order to obtain the results of Figure 6, we use the martensite phase as reference. As can be noticed, the MFI strain is irreversible in the temperature range measured due to the thermal hysteresis, but at 305 K, the material partially recovers its austenite phase.

The material presents large deformations under low magnetic fields whenever the measurement temperature is close to the MST temperature. When the transformation begins, the contraction rate is small due to the phase coexistence; however, in the region in which an abrupt deformation occurs, the sample presents a large strain for a small magnetic field change. At 297.4 K, the material presents deformation of 1.6% for a magnetic field variation of 0.25 T, but 90% of this strain is achieved within an interval of only 24 mT. The high value of MFI deformation of 2.6% is only obtained when the sample transforms completely from the cubic austenite phase to the tetragonal martensite one.

For a polycrystalline compound, it is remarkable to find such high values of deformation induced by magnetic fields below 1 T. Besides the result at 297.4 K shown above, at 298 K the martensite phase is induced by a magnetic field of only 0.68 T with a strain of 1.9%. For technological applications, such low magnetic field intensities inducing large strains is attractive due to the possibility of using smaller magnets. For comparison, a strain of 0.1% demands 4 T in the polycrystalline $Ni_{41}Co_9Mn_{31.5}Ga_{18.5}$ compound at 340 K, even without a compressive pre-strain as in our measurements.[3] Kumar et al reported that under static compressive load of 1.3 MPa, polycrystalline Mn-rich $Ni_{50}Mn_{29}Ga_{21}$ with the 7M modulated martensitic structure presents a MFI strain of 0.7% under 0.5 T at 300 K.[8] Therefore, the $Ni_2Mn_{0.7}Cu_{0.3}Ga_{0.84}Al_{0.16}$, with less than half that magnetic field and without the use of pre-strain, presents a strain a factor twice as large as that of $Ni_{50}Mn_{29}Ga_{21}$. In order to test the reproducibility of these results we repeated one of the magnetic field dependent strain measurements six times (using the same heating and cooling procedure as before). Both the critical field and the value of deformation remained the same in all the cycles.

The difference between the magnetic susceptibility of the martensite and the austenite structures is the key property to induce the structural change under a magnetic field. Therefore, it is interesting to use materials with MST or with an antiferromagnetic martensite phase.[4] In a material with MST, we need to consider that the martensite phase, which is ferromagnetic with the easy axes correlated to the misaligned twinned microstructure, has larger magnetocrystalline anisotropy energy (MAE) than the paramagnetic austenitic one, which is magnetically isotropic and cubic. Then, a higher MAE of the martensite phase imposes a barrier to the transformation from the parent phase. Whenever the temperature decreases, the enhanced interaction between the atoms in the structure results in a higher energy available to overcome the larger MAE of the low temperature martensite phase. To surpass this energy barrier under magnetic field, the magnetic interactions of the martensite phase must be higher, and therefore, the magnetic saturation of the low temperature phase must be larger. To occur at low magnetic fields, a MST depends on a high Zeeman energy required to overcome the of the martensite phase,[4] and this is the case since the Zeeman energy is proportional to the difference of magnetic saturation between the phases, it is higher in a MST because of the para-ferro transition. From the magnetization measurement shown in the Figure 5, we calculated a magnetization increase of 10 $Am^2kg^{-1}$ in a field interval of only 20 mT around 0.75 T.

Therefore, the application of a low magnetic field in the paramagnetic cubic phase at a temperature close to the MST contributes to induce the ferromagnetic phase.

### 3.5 Interplay between the MFI strain and the phase coexistence

The temperature dependence of the material MFI strain as well as the critical magnetic field for each temperature is intrinsically linked to the degree of phase coexistence, which can be evaluated by the presence of the martensite phase, as shown in Figure 7. Extrapolating the dilatometry curve for both phases allowed us to quantify the phase coexistence. The strain saturates around 2.6% corresponding to the full phase change. The critical field ($H_C$) determined from the abrupt jump in the magnetostriction measurements and plotted in the temperature vs magnetic field plane exhibits a linear tendency with $\Delta H_C/\Delta T = 1$ $TK^{-1}$. By observing the martensite phase fraction curve obtained from thermal expansion and the temperature in which the sample presents the lowest $H_C$ (0.25 T), we estimate approximately 40% of the sample to be in the martensite phase at that temperature (297.4 K). This value is in agreement with the 1.6% of deformation, which corresponds to 38.5% of temperature-induced strain before the isothermal measurement. Although the maximum MFI deformation is not reached under small fields in this polycrystalline compound, these values are still impressively large and potentially useful for application.[1,2,7,10]

## 4. Conclusion

Off-stoichiometric $Ni_2MnGa$-based Heusler alloys have been studied due to their ferromagnetic shape memory effect. The polycrystalline $Ni_2Mn_{0.7}Cu_{0.3}Ga_{0.84}Al_{0.16}$ compound studied here presents a magnetostructural transition with giant deformation of 2.6% and the possibility of induction by magnetic field. An impressively large strain of 1.6% under a low magnetic field of 0.25 T was observed at 297.4 K. The combination of a large deformation and a small magnetic field is made possible due to the transition sharpness between the paramagnetic cubic austenite and the ferromagnetic tetragonal martensite phases without special treatment of the material (for example by introduction of pores), suggesting an inexpensive route to achieving useful strain properties. The deformation values are higher than that observed in others polycrystalline FSMAs and are comparable to single crystals of Heusler alloys with non-modulated martensite phase.[4,8,27] These results motivate new studies on multifunctional alloys with abrupt magnetostructural transition as well as their technological development.

## Acknowledgements

The authors acknowledge the financial support of the CNPq (Conselho Nacional de Desenvolvimento Científico e Tecnológico), FAPERJ (Fundação de Amparo à Pesquisa do Rio de Janeiro) and CAPES (Coordenação de Aperfeiçoamento de Pessoal de Nível Superior). S. J. Stuard was supported by the Fulbright Study/Research grant.

## Contributions

A.A.M., L.E.L.S, J.F.J. and S.J.S fabricated the sample. A.A.M. and J.F.J. performed and analyzed the x-ray diffraction measurements, A.A.M., L.G., G.G.E. and A.M.G. performed the dilatometry measurements. A.A.M. and A.M.G performed and analyzed the magnetization and calorimetry measurements. A.A.M and A.M.G. wrote the manuscript. The authors discussed the results and commented on the manuscript at all stages.

# Figures

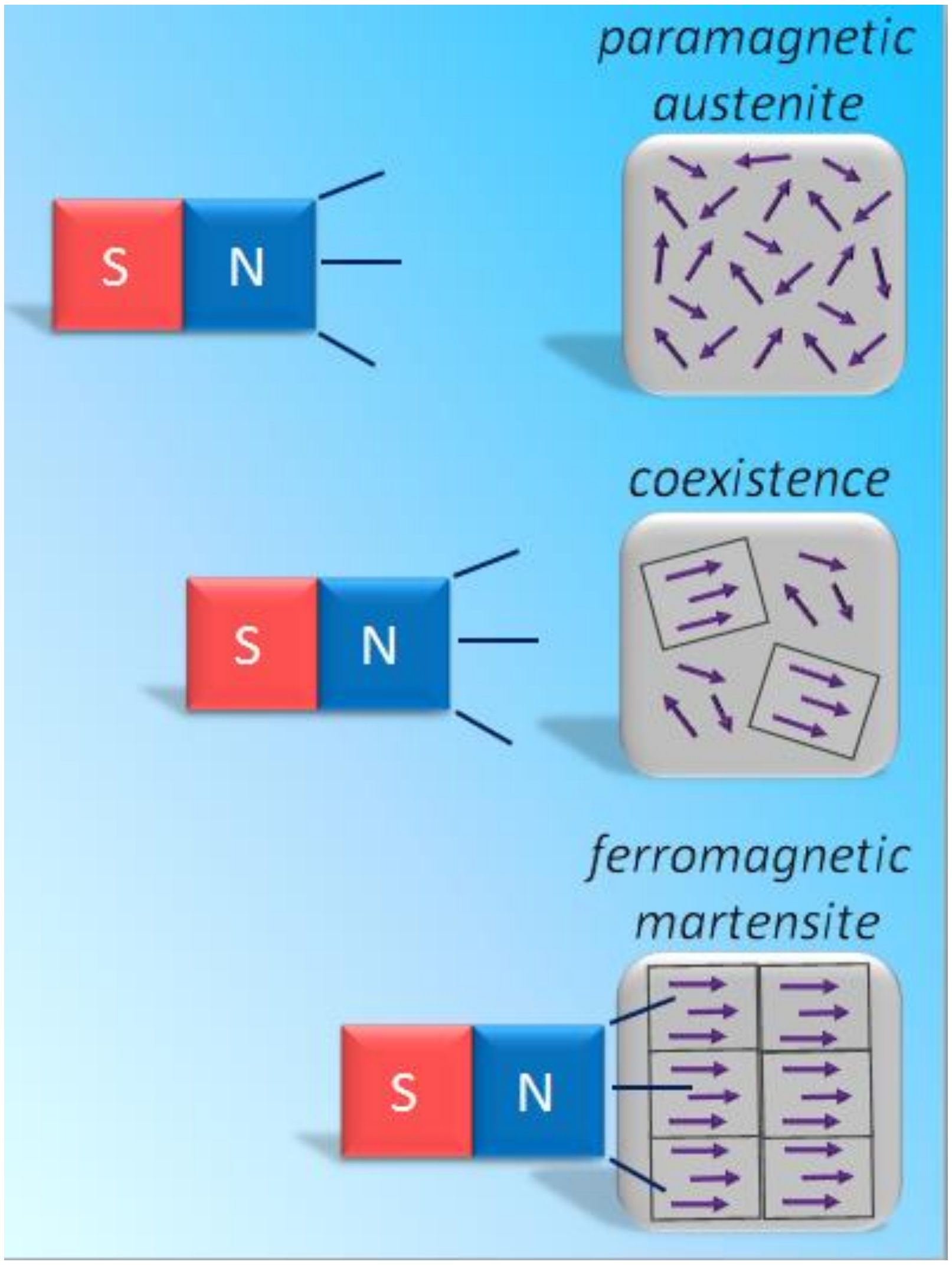

**Figure 1**. Schematic illustration of magnetostructural transition induced by magnetic field in the system with paramagnetic austenite and ferromagnetic martensite. Initially, the material is in the paramagnetic austenite phase. When the magnetic field increases, the compound presents phase coexistence due to the nucleation and growth of the ferromagnetic martensite phase. Then, at a certain intensity of magnetic field, a fully martensite induction is yielded. The arrows illustrates the volume magnetization.

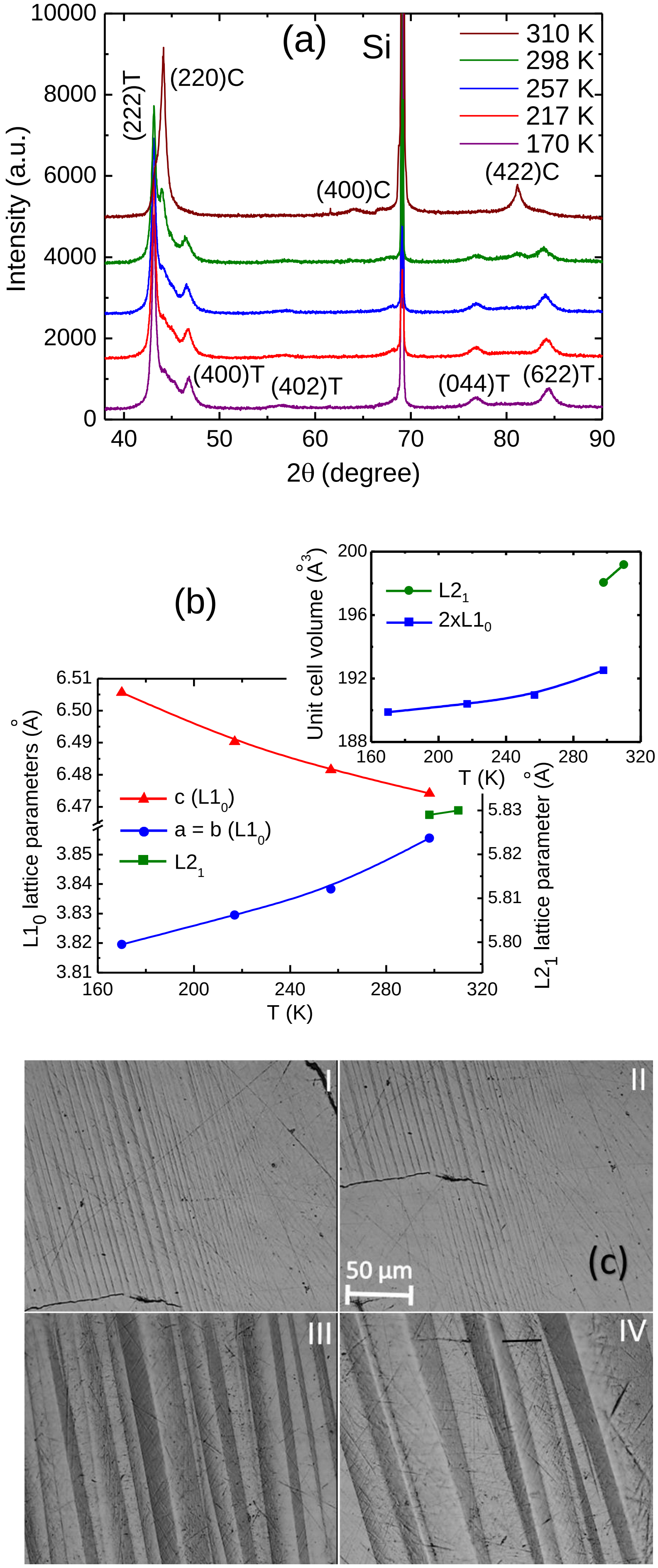

**Figure 2.** a) X-ray diffraction patters of studied alloy at different temperatures for the tetragonal martensite ($L1_0$) and cubic austenite ($L2_1$) structures. The curves are shifted vertically for better visualization. b) Optical micrographs of studied alloy obtained at T = 298 K. I, II, show the sample places with formation of martensite phase from austenite one; III and IV show the sample places with mainly martensite phase. All micrographs have the same scale. c) lattice parameters $a = b$ and $c$ of the $L1_0$-type tetragonal martensitic unit cells and $a$ of the $L2_1$-type cubic austenite unit cells a function of the temperature. Inset: Unit cell volume of the cubic and tetragonal unit cells. Since the cubic cells have twitce atoms, we ploted the volume of the tetragonal cell multiplied by 2 in order to improve the visual comparison.

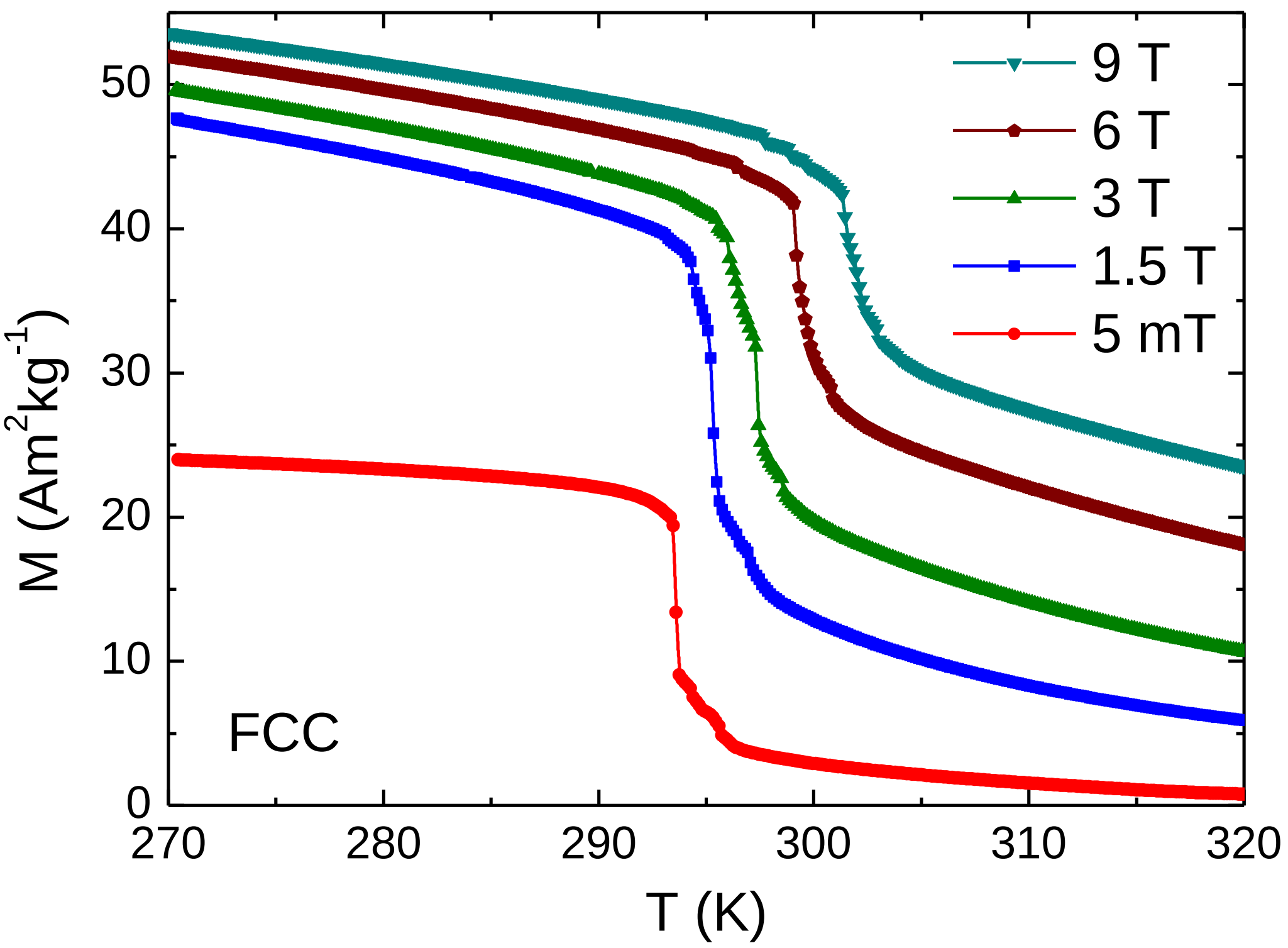

**Figure 3.** Field Cooled Cooling (FCC) magnetization under several magnetic fields. The data from 5 mT was multiplied by a factor of 20 in order to improve the visualization. The transition temperature rises when the field increases due to Zeeman energy contribution to free energy of the system.

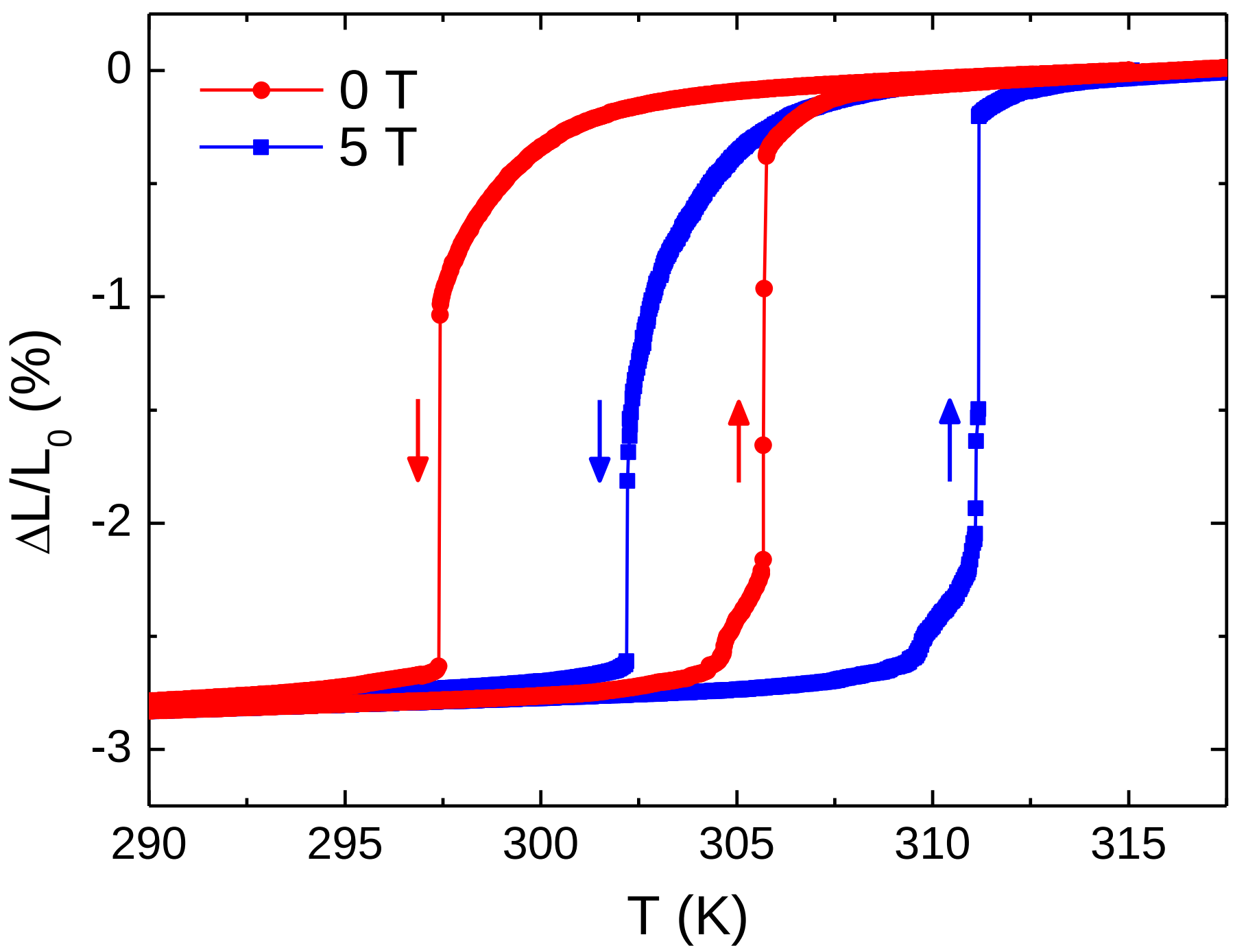

**Figure 4**. Dilatometry as a function of temperature at 0 and 5 T. The material presents large deformation of 2.6% in the MST with thermal hysteresis of around 8 K. Application of magnetic field leads to increase of the transition temperature, while keeping the strain value. The applied magnetic field was parallel to the deformation measured.

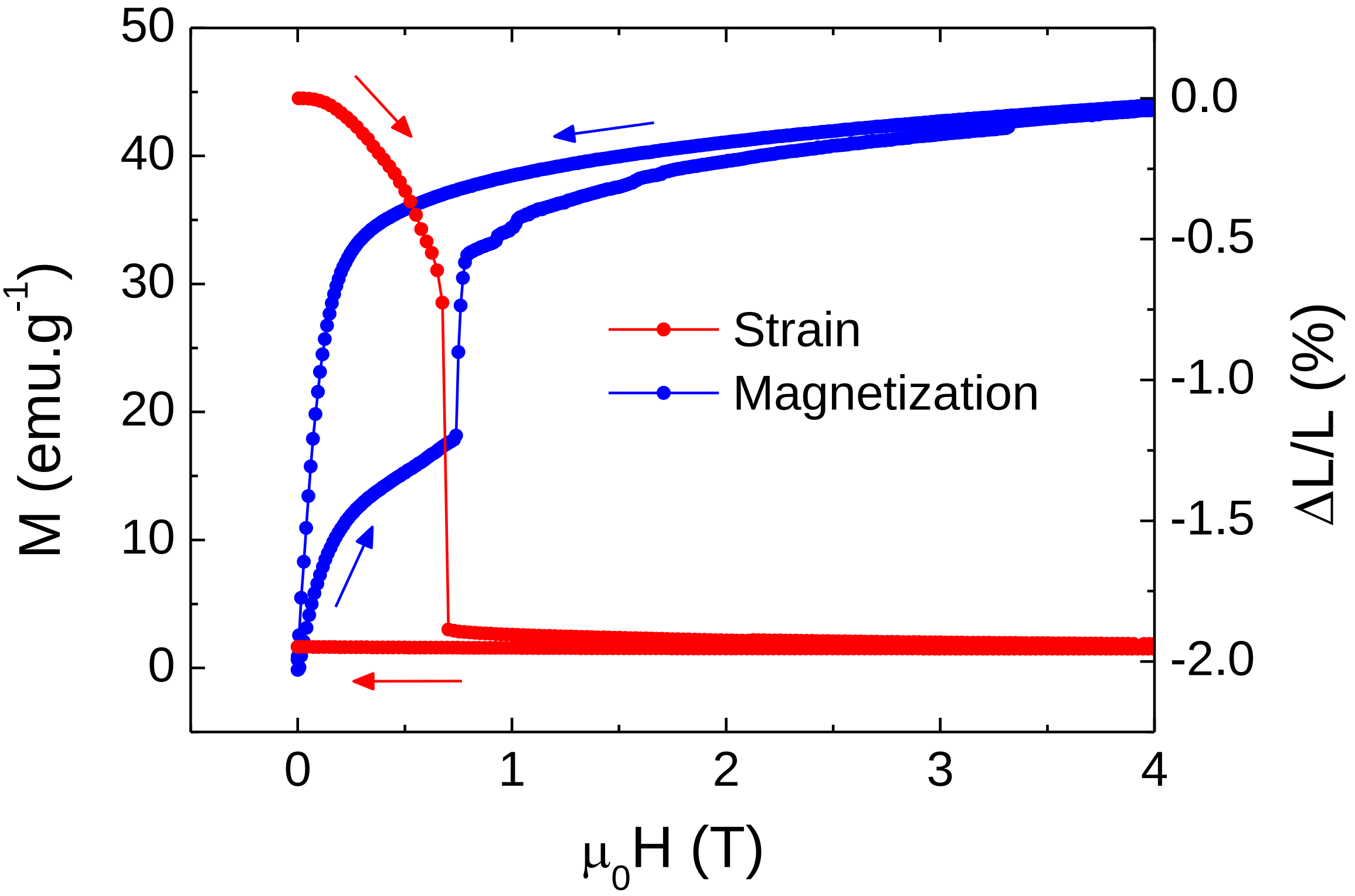

**Figure 5.** Magnetic field dependence of magnetostriction and magnetization as at 298 K. The material was cooled from temperature T = 317 K of the austenitic phase before each measurement. An abrupt behavior is noticed in the metamagnetic transition.

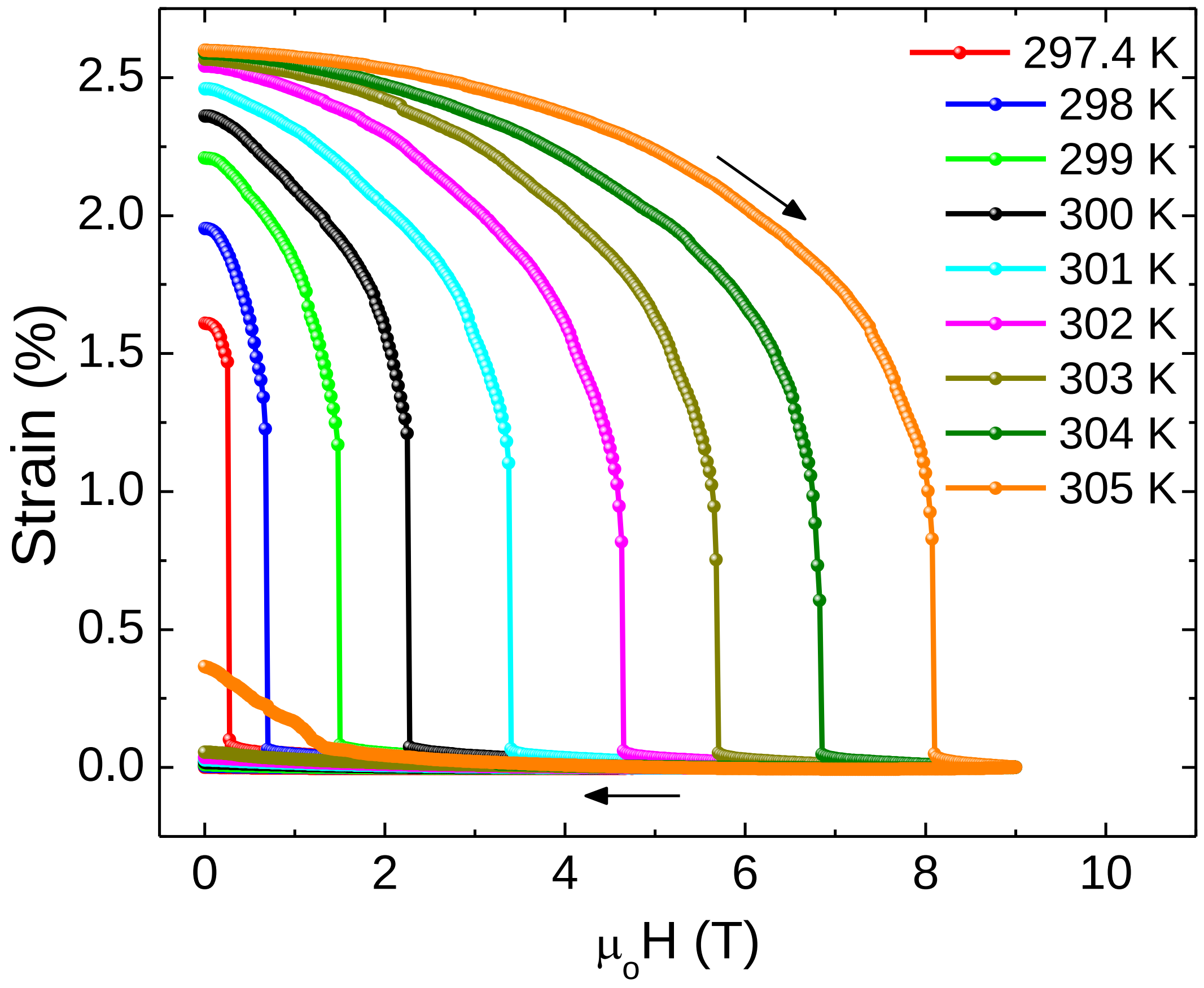

**Figure 6**. Magnetostriction measurements at several isotherms. The sample was heated to temperature T = 317 K of the austenitic phase before each isothermal measurement. High values of strain can be obtained by application of low magnetic fields due to the abrupt behavior of the transition between the paramagnetic cubic and the ferromagnetic tetragonal phases.

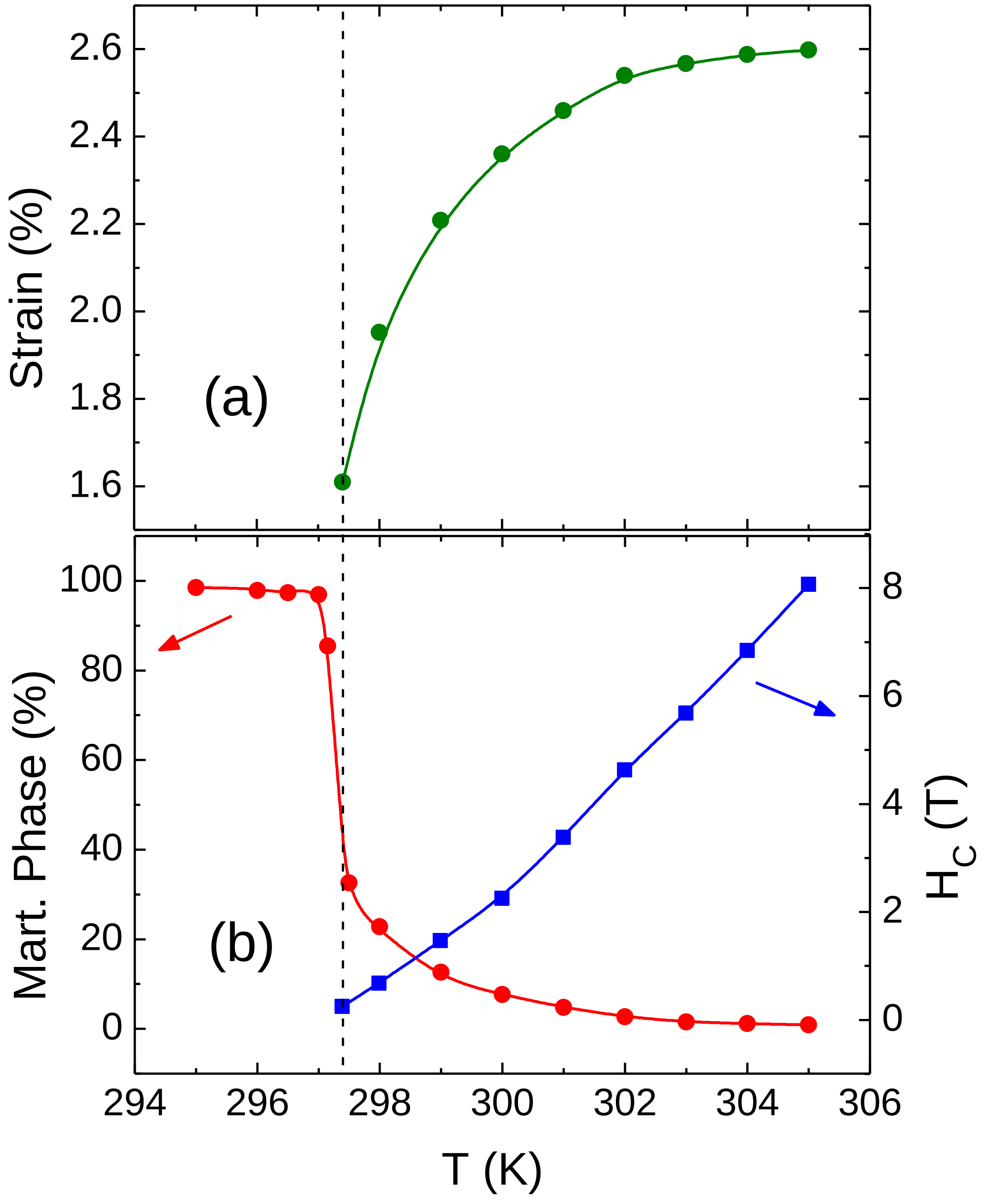

**Figure 7**. Strain in part (a) as well as martensite phase and critical magnetic field ($H_C$) in part (b) as a function of temperature. The $H_C$ (blue data) presents linear behavior, with slope close to 1 $TK^{-1}$. Gradual increase of the martensite phase is observed in the austenite phase from 305 K down to 297.4 K, where an abrupt change occurs.

**Table 1**. Temperature dependence of the lattice parameters, unit cell volume and *c*/*a*' (*a*' = $\sqrt{2}.a$) for the $Ni_2Mn_{0.7}Cu_{0.3}Ga_{0.84}Al_{0.16}$ compound. $L2_1$ is cubic and $L1_0$ is tetragonal.

| Temperature (K) | $a = b$ ($L1_0$) (Å) | $c$ ($L1_0$) (Å) | $a$ ($L2_1$) (Å) | Unit cell volume (Å) | $c/a'$ |
|---|---|---|---|---|---|
| 170 | 3.820 | 6.506 | -- | 94.94 | 1.204 |
| 217 | 3.830 | 6.490 | -- | 95.20 | 1.198 |
| 257 | 3.838 | 6.482 | -- | 95.48 | 1.194 |
| 298 | 3.856 | 6.474 | 5.829 | 96.26 ($L1_0$), 198.05 ($L2_1$) | 1.187 |
| 310 | -- | -- | 5.840 | 199.18 | -- |